\documentstyle[11pt,newpasp,twoside]{article}
\def\edcomment#1{\iffalse\marginpar{\raggedright\sl#1\/}\else\relax\fi}
\reversemarginpar

\begin{document}
\title{Brown Dwarfs as Ejected Stellar Embryos: \\
Observational Perspectives
}
 \author{Bo Reipurth}
\affil{Institute for Astronomy, University of Hawaii, 2680 Woodlawn Drive, 
Honolulu, HI 96822, USA}
\author{Cathie Clarke}
\affil{
Institute of Astronomy,
Madingley Road, Cambridge CB3 0HA, UK
}

\begin{abstract}
We discuss a scenario in which brown dwarfs are formed like stars,
except that their full collapse phases are interrupted through
dynamical interactions in small multiple systems, leading to the
ejection of the lightest member. This disintegration is a stochastic
process, often resulting in the expulsion of newborn low mass stars,
but when it occurs early enough the ejected stellar embryo will be a
substellar object. This process may be so common at early ages that a
large fraction of the ubiquitous brown dwarfs could have formed in
this manner.  Detailed gas dynamical simulations are required in order
to better understand the details of the decay of small newborn
multiple systems.  We discuss the observational consequences of the
ejection hypothesis, noting especially the importance of binaries with
brown dwarf components as an observational test. Finally, we note that
brown dwarfs that have recently been ejected may be so disturbed, by
infall from the collapsing core and also by heavy accretion from
perturbed circumstellar disks, that traditional spectral and
luminosity criteria may fail to identify their substellar nature.

\end{abstract}

\section{Introduction}

  Our understanding of the formation of brown dwarfs has long been
shrouded in controversy. Two aspects of their formation appear
problematical from a theoretical point of view. The first is that
brown dwarf masses are at least 10 times smaller than the typical
Jeans mass in star forming molecular clouds. Thus any successful
theory needs to produce regions of exceedingly high gas density ($\sim
10^7$ cm$^{-3}$ for isothermal gas at $10$K). Secondly, once
such proto-brown dwarfs have formed, it is necessary that they avoid
substantial further accretion of gas from their environment, in order
that they remain as low mass entities. 

  A variety of scenarios have been proposed that address the density
issue. Many of these involve the formation of brown dwarfs within
disks.  For example, Lin et al. (1998) suggested that the requisite
high densities would be produced during the encounter between two
massive disks, when a long tidal filament could be flung out, leading
to the subsequent formation of a brown dwarf. In loose T associations
such encounters are likely to be extremely rare, which is in contrast
to the increasing number of brown dwarfs found also in such
environments. In the simulations of Bate (2002), however, where stars
form in compact groups, such interactions are common and the majority
of brown dwarfs form in massive disks that are highly disturbed by
dynamical encounters. Li (2002), on the other hand, has shown that
brown dwarfs may also form in isolated disks if these are supported by
magnetic pressure. By contrast, Padoan \& Nordlund (2002) have argued
that brown dwarfs need not form in disks, but that sufficiently high
densities may be produced in the shock compressed regions of turbulent
flows; this process they term `turbulent fragmentation' to
distinguish it from the above scenarios in which the high densities
required to form brown dwarfs are the {\it result} of self-gravity.

 Once brown dwarfs have been produced by any of the above mechanisms,
the remaining problem is to prevent their exceeding the hydrogen
burning mass limit due to continued accretion. Recently, Whitworth \&
Zinnecker (2002) have suggested that photoionization may erode higher
mass gas cores, so that the remnant core is reduced to substellar
masses; evidently this mechanism is restricted to regions containing
photoionizing OB stars. Otherwise, there are two possibilities -
either the statistics of the density field are such that brown dwarfs
collapse in isolation, well away from the gas reservoirs that are
destined to form stars, or else, if brown dwarfs and stars form from
common gas reservoirs, the brown dwarfs must somehow be removed from
this environment.

  Reipurth \& Clarke (2001) have developed this latter line of
argument, appealing to the dynamical interactions that occur within
small N groupings in order to eject brown dwarfs from their natal gas
reservoir. The spectacular simulations of Bate (see contribution this
volume) illustrate both the propensity of molecular clouds to fragment
into small N groupings, and the way that dynamical interactions indeed
eject brown dwarfs.

  In this contribution, we review the ejected stellar embryo scenario
and its observable consequences, first considering the observational
evidence that stars are indeed formed in small N groupings. We should
stress, however, that the implication is not that all brown dwarfs
{\it must} form this way, but rather that a significant fraction can
be expected to be formed in, and ejected from, multiple systems. It is
our purpose here to set out the observational discriminants that can
be used to assess what fraction of brown dwarfs form in this way. We
emphasize that most of these discriminants must be sought in {\it
young} brown dwarfs: after a few hundred million years, the appearance
and kinematics of substellar objects depend only on mass and age, and
the particular mechanism that produced them is lost in the mist of
time.

\section{Multiplicity of the Youngest Stars}

In a detailed study of 14 driving sources of giant Herbig-Haro flows,
Reipurth (2000) found that more than 80\% are binaries, and of these
half are higher order systems. Given that such a survey can only
under-estimate the true incidence of faint companions, and if all
newborn stars go through a phase producing giant outflows, this may
then suggest that the typical star forming environment contains several
stars.

Embedded outflow sources are of the order of 10$^5$~yr old or less,
and it follows that some of these systems must decay to reach the
lower observed frequencies at later evolutionary stages. It is well
established that non-hierarchical triple systems undergo rapid
dynamical evolution and generally decay into a binary and an unbound,
escaping third member. The binary system that is formed in this
dynamical process is highly eccentric, and given that the triple
disintegration is likely to take place while the stars are still
actively accreting gas from an infalling envelope, it follows that the
circumstellar disks will interact on an orbital timescale, which will
lead to shrinkage of the orbit (e.g. Artymowicz \& Lubow 1996). These
interactions are again likely to cause cyclic variations in the
accretion rate, with consequent pulses in the outflow production, and
the giant Herbig-Haro flows may therefore represent a fossil record of
the birth and early evolution of binary systems (Reipurth
2000). Altogether, it appears that the generation of giant Herbig-Haro
flows, the birth of a binary, and the formation of brown dwarfs may
all be different aspects of a single event, namely the dissolution of
a small multiple system.

\section{Brown Dwarfs as Ejected Stellar Embryos}

  If stars form in small non-hierarchical groups (Larson 1972), then
evidence for these initial conditions is rapidly lost due to the
dynamical disintegration of such groups.  For a non-hierarchical group
containing N stars, then in the limit of small N (less than a few
10s), the timescale for the group's dissolution is roughly N crossing
times. This means that even young systems such as the Orion Nebula
Cluster, where the stellar distribution is very smooth at an age of a
few Myr (Bate et al. 1998), could easily have evolved from an ensemble
of compact few body groupings (Scally \& Clarke 2002).  Traces of
small N clustering must instead be sought in younger (Class 0)
sources (see Section~2).
 
 The reason that such groupings dissolve is primarily because, as the
system evolves, there are occasions when three stars pass close to
another (i.e. close enough that their relative orbits are strongly
perturbed; for stars with relative velocity $v$ and mass $M$, the
three stars need to pass within a distance of $ \sim GM/v^2$ of each
other). In such {\it close triple encounters}, the interchange of
energy within the system will in general cause one of the stars to
acquire more kinetic energy at the expense of hardening the relative
orbit of the other two. The repetitive action of such encounters
causes the cluster to become unbound (the total energy, in this case,
is of course conserved, and the kinetic energy of the escaping stars
is released by the formation of a binary star system). Numerous
simulations of the few body problem for a system of gravitating point
masses (e.g. van Albada 1968; Harrington 1975; Valtonen \&
Mikkola 1991; Sterzik \& Durisen 1995, 1998) have shown that the
resulting binary usually contains the two most massive cluster
members. As pointed out by McDonald \& Clarke (1993), the implication
of this result is that if stars form in this way, low mass stars and,
particularly, brown dwarfs, should be less likely to be found in
binary systems.

 In reality, the situation explored by such Nbody simulations (whereby
the stars have pre-assigned masses and interact through purely point
mass gravitational interactions) is very artificial, and the system
should instead be investigated as an initially gas dynamical one. Gas
(whether distributed or in the form of circumstellar disks) modifies
the dynamical decay process, since dissipation provides an additional
channel for the redistribution of energy in the system (Clarke \&
Pringle 1991, McDonald \& Clarke 1995).  In addition, distributed gas
(if it contains a significant fraction of the initial cluster mass)
shapes the stellar masses as it is differentially accreted on to the
stars as the cluster dissolves.  Pilot simulations have shown that
this situation is highly inequitable, and that even if the cluster
contains protostars of initially equal masses, the interplay of gas
dynamics and few body stellar dynamics produces a large dynamic range
of final stellar masses (Bonnell et al. 1997, 2001).

 Reipurth \& Clarke (2001) first explicitly linked the small N
cluster scenario to the production of brown dwarfs. They emphasized
that in the presence of gas, it is not merely the case that brown
dwarfs are ejected because they are of low mass (as is the case in
purely Nbody calculations) but that instead {\it brown dwarfs are of
low mass because they are prematurely ejected}. In this scenario, the
final mass of a star (or brown dwarf) is intimately linked to its
dynamical and accretion history and one would therefore anticipate
correlations between an object's mass and other characteristics (such
as its binarity, circumstellar environment and kinematics).  In
particular, if brown dwarfs indeed form within cores whose total mass
considerably exceeds that of a typical brown dwarf, then objects that
end up as brown dwarfs must have suffered an ejection event (or, more
probably, a sequence of close encounters culminating in one which
imparted them with positive energy).

 The typical length scale of such close encounters, $D$, may be
imprinted on the resulting brown dwarfs in several ways. Firstly, the
ejection speed is roughly related to the orbital velocity of the dwarf
during close encounter; if the typical object in the cluster has the
mass of a low mass star, then the ejection speed may be crudely
parameterized as $ 15$~$D_{AU}^{-0.5}$ [km~s$^{-1}$], where $D_{AU}$ is the
close encounter distance in AU  (Armitage \& Clarke
1997). Secondly, it is unlikely that the dwarf would be able to take
with it a reservoir of circumstellar material with a size scale
exceeding $D$. This limits both the mass and initial radius of disks
in ejected objects. Subsequent viscous evolution can cause the disk to
re-expand, but its lifetime is likely to be reduced as a result of the
encounter. Thirdly, the object cannot be ejected as a binary if $D$ is
less than or comparable to the binary separation.  Thus one would not
expect wide brown dwarf binaries (with separations greater than $\sim
D$) to be produced by this mechanism.

  Further quantification of these trends requires detailed numerical
simulations.  Recently, Bate (see contribution this volume) has placed
such interactions within the context of a star forming molecular
cloud. In these simulations (see Bate 2002, and Bate, Bonnell, \& Bromm
2002), supersonic turbulence produces a system of sheets and filaments
of compressed gas, whose further fragmentation gives rise to a number
of small N clusters.  These clusters undergo the processes described
above, although their evolution is more complex since each cluster is
no longer a `closed box' (as in previous studies) but may experience
both further infall of gas as well as additional interactions as
mini-clusters collide with each other. (Star formation activity is
centered on the filaments in this simulation, so that cluster-cluster
collisions tend to occur at the intersections of filaments as clusters
fall along the filaments).  The stars and brown dwarfs produced in the
simulation (about $20$ of each) all initially form at a mass scale of
$\sim 10$ Jupiter masses, which is the minimum Jeans mass allowed by
the equation of state employed. Their final masses, which roughly
conform to the observed IMF in the stellar regime, are thus entirely
set by the process of competitive accretion within a small N cluster
environment.

 A complementary investigation led by Delgado-Donate (see contribution
this volume) has instead focused on multiple random realizations of
small N clusters as isolated hydrodynamic systems. By considering
isolated clusters, these simulations obviously cannot follow the
cluster-cluster interactions that are a feature of the whole cloud
simulations of Bate.  In return, the computational economy effected by
not modeling the larger cloud environment (about a factor of $6$ in
terms of CPU hours per star produced) allows one to obtain
statistically significant numbers of stars for each initial condition
and thus to explore the connection between initial conditions and
resultant stellar properties. Furthermore, in contrast to Bate's
simulations, these isolated simulations have been integrated as Nbody
systems following the accretion of the gas, until they have decayed
into stable hierarchical multiples and single stars. In this way,
these simulations can shed some light on the properties of wide
binaries, which remain as non-hierarchical multiples in Bate's
simulations.  Preliminary results from this investigation suggest that
the statistical properties of the resulting stars and binaries are not
unduly sensitive to the power spectrum of perturbations employed in
the initial conditions; this suggests that the combination of
fragmentation and competitive accretion may produce a rather robust
set of stellar parameters, weakly dependent on the detailed
environment.

\section{Observational Consequences of the Ejection Model}

In the following we summarize a number of observations, which may cast
light on the validity of the ejection model. 

{\em Brown Dwarfs near Class 0 Sources.}

The bulk of the mass of a newborn star is accumulated during its
embedded phase, which lasts $\sim$10$^5$ yr. Such nascent objects are
seen as Class~0 and Class~I sources. The strong outflow activity of
Class~0 sources suggest that it is during this evolutionary stage that
a star builds up most of its mass. It follows that if the accretion
process for a stellar embryo should be interrupted early enough to
limit its mass to substellar values, the dynamical interactions
leading to break-up most likely occur during the Class~0 phase. How
long time a newborn brown dwarf will be observable in the vicinity of
an embedded source will then depend primarily on the ejection
velocity. Assume that a newborn brown dwarf is observed at a time {\em
t} (years) after ejection, that it is moving with a space velocity
{\em v} (km per sec) at an angle {\em $\alpha$} to the line of sight,
and that it is located in a star forming region at a distance {\em d}
(parsecs). The projected separation {\em s} in arcseconds from the
embedded source will then be $ s = 0.21vtd^{-1} sin \alpha.$ As an
example, consider a newborn brown dwarf in a nearby cloud ($d \sim
130~pc$) moving at an angle of 60$^o$ to the line of sight with a space
velocity of 1 km~s$^{-1}$. After 10$^5$ yr, when the embedded phase of
the original Class~0 object is about to end, the brown dwarf is
already 140 arcsec away, with no obvious connection to its site of
birth. Furthermore, half of all ejected brown dwarfs will move into
the cloud in which they were born, appearing as very weak infrared
sources, thus making their detection more difficult. Altogether, if
sufficiently many Class~0 and perhaps Class~I sources are surveyed, a
number of newborn brown dwarfs could be found. A likely brown dwarf
candidate was found 4 arcsec from the deeply embedded driving source
of the HH~111 jet (Reipurth et al. 1999).

{\em Kinematics of Brown Dwarfs.}

If brown dwarfs are ejected stellar embryos, they should carry
kinematic signatures reflecting their origin in small multiple
systems.  The velocity dispersion of young stars is generally assumed
to reflect the turbulent velocity of the gas out of which they formed,
that is, their velocities are expected to be of the order of a km per
second. If all stars are formed in small multiple systems, as
suggested by Larson (1972), then an additional velocity component
comes from the velocities attained when the cluster dissolves and the
members drift apart. The velocity that an ejected member acquires is
comparable to the orbital speed attained at pericenter in the close
triple encounter. Realistic numerical simulations including gas
dynamics have been carried out by Delgado-Donate and Clarke (see their
contribution in this volume), and they conclude that typical
velocities of brown dwarfs are mostly less than 2 km~s$^{-1}$.
One-dimensional velocities (radial or tangential) are correspondingly
smaller.  This is so close to the general turbulent velocity
dispersion in molecular clouds that it will be difficult to test the
ejection scenario kinematically.  Further, Delgado-Donate and Clarke
find that there is {\em no} dependence of the emergent speed on
stellar mass amongst single ejected stars, although they find (in
common with Sterzik \& Durisen 1998) that binary systems emerge with
much lower velocities.  Since the binary fraction is seemingly higher
among stars than brown dwarfs, this would translate into an effective
mass dependence of the emergent velocity.

Most brown dwarfs are, like stars, born within larger scale clusters.
With the velocities quoted above, most stars and brown dwarfs would be
retained within typical open clusters, but, after a crossing time or
so, the distribution of brown dwarfs would be more extended due to
their higher initial velocity dispersion. However, after about $10^8$
years, the velocities of cluster members are dominated by dynamical
relaxation, which causes the brown dwarf distribution to be somewhat
more distended than the stars (de la Fuente Marcos \& de la Fuente
Marcos 2000, Adams et al. 2002), irrespective of the initial
velocities with which they are injected into the cluster (Moraux et
al., in prep.).  A proper test of the ejection scenario is made only
by finding a halo of brown dwarfs around clusters small enough and/or
young enough that relaxation has not yet dominated their kinematics,
but the size of the expected effect may be marginal. Finally, we note
that kinematics of isolated old brown dwarfs no longer reflect
conditions at the time of birth, since most will have evaporated from
clusters.

{\em Infrared Excesses of Brown Dwarfs.}

The members of newborn multiple systems are likely to be surrounded by
substantial amounts of circumstellar matter. However, as a prelude to
dynamical ejection, a brown dwarf must go through a close triple
encounter, which will prune its circumstellar disk. Freefloating brown
dwarfs may therefore have smaller reservoirs from which they can
accrete and form planets. Armitage \& Clarke (1997) showed that disk
truncation in T~Tauri stars could promote the rapid decline of
classical T~Tauri characteristics and, in analogy, we would therefore
expect that ejected brown dwarfs will display the signatures of infall
and outflow for a more limited time. But if encountered at an early
enough age, a brown dwarf should show just the same indicators of disk
accretion as any young star. In particular it should be stressed that
the detection of near-infrared excesses in brown dwarfs, and thus the
presence of circumstellar disk material, is not at variance with the
ejection hypothesis. Excesses at near-infrared wavelengths refer to
the inner edges of disks at distances of only a few stellar radii from
the central object, regions which are unaffected by the disk
truncation. If a brown dwarf disk is truncated at a radius of, say,
20~AU (as is typical in the simulations of Bate 2002), then the
observed spectral energy distribution is affected only at wavelengths
longer than about 40 $\mu$m (Natta \& Testi 2001; Natta et al. 2002). The
near- and mid-infrared energy distributions are, however, sensitively
dependent on the geometry of the disk, in particular the flaring
angle, about which very little is known for brown dwarfs (see also
Testi et al. in this volume).

{\em The Brown Dwarf Desert.}

The ``brown dwarf desert'' refers to the long known fact that brown
dwarfs are only rarely found as close (less than 3~AU) companions to
low mass stars (on these scales the incidence of brown dwarf
companions to stars with masses of 0.5~M$_\odot$ or more does not
exceed 1\%).  Recently, however, Gizis et al. (2001) have demonstrated
that at large separations (greater than 1000~AU) one commonly finds
brown dwarf companions to normal stars.

The ejection hypothesis readily explains the absence of brown dwarfs
close to normal stars, because of the combination of two effects. As
is well known from pure N-body simulations of small N clusters,
exchange interactions tend to substitute more massive cluster members
into binaries that temporarily contain only low mass components. But,
as shown in gas dynamical simulations, continued accretion tends to
equalize binary mass ratios (Artymowicz 1983, Bate 2000), so that a
brown dwarf companion may not remain as such unless the binary is
swiftly removed from the reservoir of gas that feeds it. Only under
special circumstances (see Reipurth \& Clarke 2001) will dynamical
interactions directly produce a brown dwarf as a close companion to a
star (but it may arise through subsequent orbital evolution if the two
components have circumstellar material that interacts).  On the other
hand, the ejection hypothesis readily explains the presence of brown
dwarfs as distant companions to stars, simply because not all
ejections lead to an unbound system.

Of course, it is possible that the brown dwarf desert may not be a
remnant of the formation process, but could result from orbital
migration (Armitage \& Bonnell 2002).

{\em Brown Dwarf Binaries.}

An increasing number of brown dwarfs have been found to be binaries
(e.g. Mart\'\i n, Brandner, \& Basri 1999).  The distribution of
separations in brown dwarf binaries contains important memories of
their formation, which any theory of brown dwarf formation must be
able to reproduce. It is notable that, up to now, only rather close
brown dwarf binaries have been discovered. Pairs with separations of
many hundreds of AU are not seen, although the observing techniques
have no obvious bias against their detection. This is equally true for
brown dwarfs in clusters and in the field, perhaps suggesting that we
are not seeing an effect of their environment, but a result of their
formation process.

From the perspective of the ejection hypothesis, there are two aspects
of this problem: the formation of binary embryos, and the survival of
the nascent binary during its ejection. The latter clearly militates
against wide binaries, since they would be torn apart in a close
triple encounter, thus being consistent with the tentative absence of
observations of wide brown dwarf pairs.

The formation of binary embryos is more problematical. McDonald \&
Clarke (1993) and Sterzik \& Durisen (1998) showed that for purely
point-mass dynamical interactions, binaries made up of two brown
dwarfs would essentially never be formed. However, stellar embryos in
small multiple systems are likely to be associated with substantial
circumstellar material, offering a source of dissipation that can
create such low mass pairs.  A further problem, however, is that such
pairs do not generally remain as brown dwarfs but instead accrete over
the hydrogen burning mass limit. As a result, brown dwarf binaries are
rather rare in the simulations of Bate (2002) and Delgado-Donate \&
Clarke (see this volume).

{\em Binarity of Stellar Primaries.}

Evidently, for a stellar embryo to be ejected as a brown dwarf
requires the presence of at least three bodies that interact. When the
third member is ejected, either into a distant but bound orbit or as
an unbound object, the two remaining components are bound more closely
in a harder binary. It is of great interest to examine the primaries
in wide binary systems with a brown dwarf companion, since the
ejection hypothesis posits that such primaries should themselves be
close binaries. Depending on the distance to the object, such binaries
may be resolved with adaptive optics techniques, or with spectroscopic
techniques.

For close binaries with a low mass primary and a brown dwarf companion
there is the possibility that the entire binary was ejected, just as
brown dwarf pairs can be ejected, but current simulations suggest that
this is not commonly happening. In such cases the primary is obviously
not expected to be itself a binary.
Clearly the study of the binarity of primaries is likely to be an
important test of the ejection hypothesis.  For a more detailed
discussion, see Sterzik \& Durisen in this volume.

{\em The Rotation Rates of Brown Dwarfs.}

If disks around brown dwarfs are truncated by the ejection process,
then this may affect the rotation rate of young brown dwarfs, since
magnetic coupling to an accretion disk is the most popular
mechanism for explaining the relatively modest rotation rates
of young stars (e.g. K\"onigl 1991, Cameron \& Campbell 1993, Armitage
\& Clarke 1996). It is not, however, clear whether the interactions
that we are positing here (within $\sim 20$ AU) are close enough to
release the proto-dwarf from its disk brake. Detailed simulations
of magnetic braking in stars suggest that the disk brake only
becomes ineffective at very low disk masses (Armitage
\& Clarke 1996). The value of this
critical disk mass depends on the moment of inertia of the star
and is higher at early times when the star is more extended. Thus,
as discussed by Clarke \& Bouvier (2000), if disk stripping is
to have any effect on stellar rotation, it must occur at early
times, which is certainly an expectation in the dynamical decay
scenario considered here. In conclusion, the sign of the possible
effect is clear - i.e. that objects whose disks are pruned would
tend to rotate faster. What is less clear is whether the interactions
we expect in mini-clusters would be close enough to make a
substantial difference to the rotation rates of young
brown dwarfs.

\section{What do Newborn Brown Dwarfs look like?}

It follows from the preceding discussion that if brown dwarfs are
ejected stellar embryos we should, at least in principle, be able to
find brown dwarfs near Class~0 sources which are extremely young, even
less than 10$^5$ yr old. From an observational point of view it may,
however, not be so clear what we should be looking for to identify
such newborn brown dwarfs. Models for relatively old brown dwarfs are
in excellent agreement with observations. But, as stressed by Baraffe
et al. (2002), for young objects the models become very sensitive to
the initial conditions, e.g. the initial radius. And for ages less
than about 1 Myr the input conditions are still so important for the
models that they cannot be meaningfully compared to observations (see
also Baraffe et al. in this volume).

So what do we look for when we try to find the youngest brown dwarfs?
The boundary between stellar and substellar objects at the age of
nearby young associations is at a spectral type around M6, so we may
expect to see strong molecular bands. However, if a stellar embryo has
very recently been exposed to heavy infall, its outermost layers may
be disturbed and heated at least for about 10$^4$ years after
ejection, which is the Kelvin-Helmholtz timescale for young brown
dwarfs. During this period the surface temperatures are possibly high
enough to dissociate molecules. Also, even though the outer parts of
the circumstellar disk of an ejected brown dwarf will be truncated,
the inner parts are likely to be extremely perturbed by the close
triple encounter (e.g. Clarke \& Pringle 1991), most likely leading to
active accretion.  So although we expect newborn brown dwarfs to have
very low luminosities, the observed luminosities may be increased
through contributions from accretion.  Spectroscopically, we may
expect that the very youngest brown dwarfs are heavily veiled objects,
perhaps obscuring all absorption features, and only showing strong
H$\alpha$ emission. And if, as seems likely, the stellar embryos also
have magnetic fields, we may expect the very youngest and heavily
accreting brown dwarfs to have small jets, which may be visible in
forbidden lines and as thermal radio jets. These are, however,
signatures of youth commonly found among very young low mass stars, so
we may find that identifying newborn brown dwarfs as bona fide
substellar objects could be a difficult task.

\acknowledgments

We are grateful to Isabel Baraffe, Matthew Bate, Alan Boss,
Eduardo Delgado-Donate, and Chris Tout for useful discussions.

\end{document}